\documentclass[paper=a4, fontsize=11pt]{article}
\usepackage[sc]{mathpazo} 
\usepackage{amsmath}
\usepackage{amsthm}
\usepackage{cite} 
\usepackage[a4paper, total={6.5in, 9in}]{geometry}
\usepackage{tabularx, booktabs}
\usepackage{hyperref}
\usepackage{comment}
\usepackage{xcolor}
\usepackage{graphicx}
\usepackage{caption}
\usepackage{subcaption}
\usepackage{bm} 
\usepackage[thinc]{esdiff}

\newcommand{\probP}{\text{I\kern-0.15em P}}

\title{Multigraph reconstruction via nonlinear random walk}

\author{Jean-François de Kemmeter$^{1}$ and Timoteo Carletti$^{1}$}

\date{%
$^1$Department of Mathematics and Namur Institute for Complex Systems, naXys, University of Namur, rue Graf\'e 2, B 5000 Namur, Belgium}

\begin{document}

	\maketitle
	
	\begin{abstract}
		
		Over the last few years, network science has proved to be useful in modeling a variety of complex systems, composed of a large number of interconnected units. The intricate pattern of interactions often allows the system to achieve complex tasks, such as synchronization or collective motions. In this regard, the interplay between network structure and dynamics has long been recognized as a cornerstone of network science. Among dynamical processes, random walks are undoubtedly among the most studied stochastic processes. While traditionally, the random walkers are assumed to be independent, this assumption breaks down if nodes are endowed with a finite carrying capacity, a feature shared by many real-life systems. Recently, a class of nonlinear diffusion processes accounting for the finite carrying capacities of the nodes was introduced. The stationary nodes densities were shown to be nonlinearly correlated with the nodes degrees, allowing to uncover the network structure by performing  a few measurements of the stationary density at the level of a single arbitrary node and by solving an inverse problem. In this work, we extend this class of nonlinear diffusion processes to the case of multigraphs, in which links between nodes carry distinct attributes. Assuming the knowledge of the pattern of interactions associated with one type of links, we show how the degree distribution of the whole multigraph can be reconstructed. The effectiveness of the reconstruction algorithm is demonstrated through simulations on various multigraph topologies.
		
	\end{abstract}
	
	\section{Introduction}
	We are surrounded by networks~\cite{Newman2010,Barabasi2016}. Physics~\cite{AB2002}, economy~\cite{Jackson2008}, biology~\cite{BGL2011} and sociology~\cite{WF1994}, are few but relevant research domains that can be studied in the vast realm of network science. Notwithstanding the differences and the peculiarities of each domain, scholars have been able to provide a transversal and unified description of those networked systems {in which} individual units (e.g., chemicals, bits of information, nutrient{s}) belonging to a certain population may move across network nodes by using the intricate web of available links, the latter defining the structure of a complex network. Those processes eventually giv{e} rise to the emergence of self-organized complex patterns~\cite{NP1977,PSV2010,BLMCH2006,BBV2008}, the latter depending on the network architecture. Hence, the bridge between network structure and dynamics can be explored by studying the evolution of simple processes; a stereotypical example is provided by a random walk, i.e., a simple model of diffusion, where basic units jump from node to node according to some microscopic rules and using available links.
	
Random walks play a central role in different fields of science~\cite{Balescu1997,bAH2000,KS2011}; once dealing with networks, a central research theme is devoted to studying the relation between patterns of diffusion and network structure~\cite{MPL2017}, with relevant applications to centrality measures related to walkers density~\cite{BP1998,LR2012}, e.g., Page rank, or community detection methods based on time of visit, e.g., Markov stability~\cite{RB2008,DYB2010,LDB2014}. In those applications, walkers are assumed to be independent from each other, a condition that can be safely assumed to hold true once no competition for the available space is at play. To overcome this limitation, a novel class of nonlinear diffusion processes on networks has been recently studied, accounting for the finite carrying capacities of nodes~\cite{ACDPFP2018,CAFL2020}. Such processes incorporate the limited available space within the nodes, a feature of most real-life networks. Take for instance the case of an ecological network in which nodes would model local habitat patches. The latter certainly cannot accommodate for an infinite number of individuals due to competition for limited resources. In other words, setting a constraint on the maximal number of agents allowed to sit on a node, naturally induces correlations between the random walkers. As a consequence, the stationary node densities are no longer simply proportional to the node degree~\cite{dKCA2022}. In some cases though, an explicit expression for the stationary density can still be worked out and thus, by taking advantage of the mass conservation, the degree distribution of a graph can be inferred from a few measurements performed at a \textit{single} arbitrarily selected node \cite{ACDPFP2018}. 

The purpose of this work is to extend the latter approach to the case of multigraph networks, where two nodes can be connected by several links, each one associated with a different attribute. Examples include air transport networks in which air routes (links) between airports (nodes) carry as attribute the name of the operating company, or social networks in which individuals (nodes) interact through several platforms (links). Assuming a finite set of distinct attributes, the corresponding non-simple networks can be represented as edge-coloured multigraphs - multigraphs for short -, each color mapped bijectively to a given value of the attribute. 

In several relevant applications, the network structure is only partially known. This might be the case of social networks in which only a fraction of relationships is available or protein-protein interaction networks in which the identification of interactions is costly \cite{lei2013novel}. Methods able to uncover the network structure from partial knowledge result thus important. With this regard, link prediction has attracted increasing attention these last years. The purpose of link prediction is to compute the likelihood of a link between two nodes, based on observed links and nodes' attributes. The latter might be structural properties such as their degree or external information, such as age or hobby if nodes refer to individuals. Various algorithms have been designed for performing link prediction, among which similarity-based algorithms and probabilistic models \cite{MBC2016,DASSA2020}. Some other models take advantage of the interplay between network structure and dynamics. In particular, random walks allow for the identification of central nodes and the detection of communities, groups of nodes more tightly connected among them than with nodes outside the community \cite{LDB2014}. 

In this work, we restrict our attention to multigraphs with two distinct attributes, i.e., two colors. We first extend the theory of non-linear diffusion process considered in~\cite{ACDPFP2018,CAFL2020} to the case of multigraphs, by assuming nodes of the latter to be endowed with a limiting carrying capacity. We explicitly determined the stationary node density and by using the fact that the latter at the level of a single node depends on the degree distribution of the whole multigraph, we are able to extend the reconstruction scheme proposed in~\cite{ACDPFP2018} to the case of multigraphs. We investigate various multigraph topologies, assessing in each case, the accuracy of the reconstruction procedure. More precisely we consider multigraphs where the known and unknown layers are selected among an Erd\H{o}s-R\'enyi graph, a Watts-Strogatz network or a scale-free one; then we reconstruct the first and second degree momenta of the unknown layer and we compare them with the ground truth as a function of the main model parameter, i.e., the probability to have a link among two nodes in the Erd\H{o}s-R\'enyi graph, the rewiring probability in the case of Watts-Strogatz and the power law exponent for the scale-free case. As a general conclusion we can assert that with local and partial information we are able to reconstruct quite well the degree probability distribution and the degree momenta, the first momentum being generally closer to the right value than the second momentum. We can also claim that our results do not suggest any strong impact of the known layer on the reconstruction of the second, let us however stress that this can be a consequence of the made assumption of absence of correlations among node degrees. In the case the unknown network is an Erd\H{o}s-R\'enyi graph we observe a degradation of the results as it becomes more sparse and in particular nodes with degree zero appear. This observation is also confirmed in the study of the real mutligraphs we carried out, where one network possesses nodes with zero degree, the whole multigraph being still connected thanks to the second network.

Moreover, still starting from the exact formula for the stationary node density we proposed a second method to reconstruct node degree momenta based on the assumption of diluted walkers density. The method requires less information than the previous one and still, it is able to reconstruct the first two momenta with a quite good precision.

In conclusion in this work we have thus strengthen the link existing between network topology and dynamical systems defined on top of networks, by proposing a model capable to infer global structural information from a multigraph with a limited amount of information, e.g., knowing the degree distribution for a single layer and measuring the random walk frequency of transit on a single node, designing in this way further studies of links prediction.

The paper is organized as follows. In Section \ref{sec:model}, we introduce the non-linear diffusion process  and derive the analytical formula for nodes' stationary densities. In Section \ref{sec:reconstruction}, we explain how the multigraph structure can be recovered as the solution of an inverse problem {based on the knowledge of the stationary node density at a single arbitrary node}. In Section \ref{sec:robustness}, we test the algorithm on various multigraph topologies and assess the robustness of the reconstruction. We then conclude and give some perspectives in Section \ref{sec:conclusion}. 
	
	\section{The model}\label{sec:model}
\begin{figure}[htb!]
		\centering
		\begin{minipage}[c]{0.3\textwidth}
			\centering
			\includegraphics[width=1.1\textwidth]{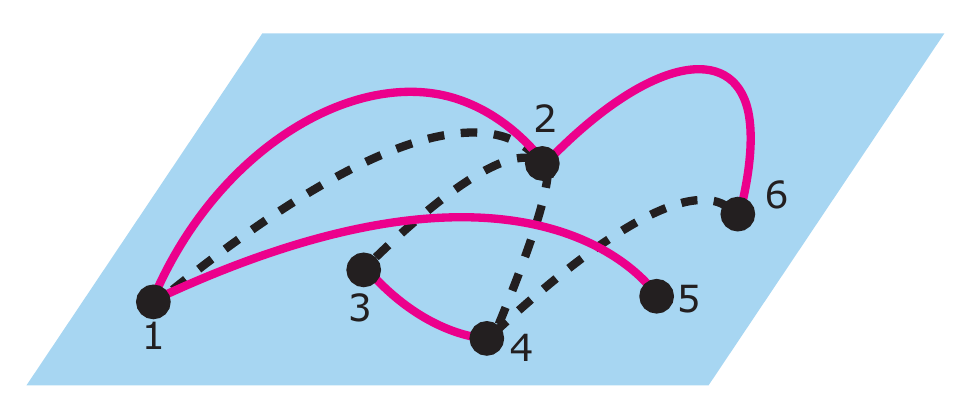}
		\end{minipage}
		\begin{minipage}[c]{0.3\textwidth}
			\centering
			$ A^{(1)} = 
			\begin{bmatrix}
				0 & 1 & 0 & 0 & 1 & 0\\
				1 & 0 & 0 & 0 & 0 & 1\\
				0 & 0 & 0 & 1 & 0 & 0\\
				0 & 0 & 1 & 0 & 0 & 0\\
				1 & 0 & 0 & 0 & 0 & 0\\
				0 & 1 & 0 & 0 & 0 & 0
			\end{bmatrix}
			$
		\end{minipage}
		\begin{minipage}[c]{0.3\textwidth}
			\centering
			$ A^{(2)} =
			\begin{bmatrix}
			0 & 1 & 0 & 0 & 0 & 0\\
			1 & 0 & 1 & 1 & 0 & 0\\
			0 & 1 & 0 & 0 & 0 & 0\\
			0 & 1 & 0 & 0 & 0 & 1\\
			0 & 0 & 0 & 0 & 0 & 0\\
			0 & 0 & 0 & 1 & 0 & 0
			\end{bmatrix}
			$
		\end{minipage}
		\caption{Multigraph with two types of links, distinguished in plain magenta and dashed black. The adjacency matrices $A^{(1)}$ and $A^{(2)}$ of the subgraphs $G_1$ and $G_2$ are indicated.}
		\label{fig:multigraph}
	\end{figure}

	In this section, we introduce the nonlinear diffusion process evolving on a multigraph. The derivation is obtained from a microscopic formulation of the dynamics and builds on previous works~\cite{ACDPFP2018,CAFL2020}.
	We start by considering a set $V$ of $\Omega$ nodes and a set $E$ of edges (see Fig. \ref{fig:multigraph} for a cartoon example). Links between nodes are distinguished by some attribute $\mathcal{A}$ which we assume to take only two distinct values, say $\mathcal{A}\in \{1,2\}$. We will denote by $E_1$ (respectively $E_2$) the subset of edges associated with the attribute $1$ (respectively $2$). The multigraph $G=(V,E)$ is then obtained as the union of the two subgraphs $G_1=(V,E_1)$ and $G_2=(V,E_2)$. The structure of each of the two subgraphs is encoded by two adjacency matrices, $A^{(1)}$ and $A^{(2)}$, with $A_{ij}^{(\ell)} = 1$ if nodes $i$ and $j$ are connected through a link with attribute $\mathcal{A}=\ell$.
	
	Given some $\beta \in (0,1)$, we then assume a set of $\lfloor \beta N \Omega \rfloor$ agents, also named walkers, to diffuse across the network under the constraint that at most $N$ of them are found simultaneously in any given node. The parameter $\beta$ thus measures the density of walkers within the network. The state of the system at time $t$ may be recorded as a vector $\mathbf{n}(t):={(n_1(t),n_2(t),\cdots,n_\Omega(t))}^\top$ with  $0\leq n_i(t) \leq N$ being the number of walkers found in node $i$ at time $t$. Denoting by $\probP{}(\mathbf{n},t)$ the probability that the system is in state $\mathbf{n}$ at time $t$, the latter obeys the \textit{M-equation}:
	\begin{equation}
	\diff{\probP{}(\mathbf{n},t)}{t} 
	=
	\sum_{\mathbf{n'} \neq \mathbf{n}} 
		\left \{
		T(\mathbf{n} \vert \mathbf{n'})\probP{}(\mathbf{n'},t)
		-T(\mathbf{n'} \vert \mathbf{n})\probP{}(\mathbf{n},t)
		 \right\},
	\end{equation}
	being $T(\mathbf{n'}\vert\mathbf{n})$ the transition probability from state $\mathbf{n}$ to state $\mathbf{n'}$. As any node may host from $0$ to $N$ agents at a given time, the terms $T(\mathbf{n'}\vert\mathbf{n})$ may be seen as the entries of a huge matrix of size $(N+1)^\Omega$. The matrix is essentially sparse as during the infinitesimal time interval $[t,t+dt]$, at most one random walker will have moved from a node to one of its neighbors. Assuming the departure node, say $i$, to contain $n_i$ walkers and the arrival node, say $j$, to contain $n_j$ ($\leq N-1$) walkers, the corresponding transition probability is written as $T(n_i-1,n_j+1\vert n_i,n_j)$. To write down the latter, let us first notice that the probability for the walker sitting at node $i$ to choose, among all the available connections incident to that node, a link leading to node $j$, is given by $\frac{A_{ij}^{(1)}+A_{ij}^{(2)}}{k_i}$, with $k_i=\sum_{j}\big\{A_{ij}^{(1)}+A_{ij}^{(2)} \big\}$ the number of connections incident to node $i$, regardless of their corresponding attributes. The probability for the walker to settle onto the arrival node is then modulated by the available space within node $j$: the more crowded the node $j$, the less likely the transition. Hence, the overall transition probability reads
	\begin{equation}
	T(n_i-1,n_j+1\vert n_i,n_j) = \frac{A_{ij}^{(1)}+A_{ij}^{(2)}}{k_i} \frac{n_i}{N} g\Big(\frac{n_j}{N}\Big),
	\end{equation} 
	where the function $g$ takes into account the available space within the arrival node. Let us recall that the case of uncorrelated walkers is recovered with $g(x)=1$. In contrast, we here impose $g(1)=0$ to forbid a walker to hop onto a fully occupied node. A common choice is to take $g(x)=1-x$. To make further progress in the analytical treatment, we now assume the nodes carrying capacities $N$ to be large enough and introduce the continuous variable, also referred to as ``node density'', $\rho_i(t) := \lim_{N\rightarrow + \infty}  \frac{\langle n_i(t) \rangle}{N}$, where $\langle \cdot \rangle$ denotes the average over stochastic realizations of the system. As shown in Appendix~\ref{app:ME2ODE}, the time evolution of nodes densities is governed by the following system of ordinary differential equations.
	\begin{equation}
	\diff{\rho_i}{t} = \sum_{j} 
	\left\{
	\rho_j\frac{A_{ji}^{(1)}+A_{ji}^{(2)}}{k_j}g(\rho_i)
	-\rho_i\frac{A_{ij}^{(1)}+A_{ij}^{(2)}}{k_i}g(\rho_j)
	\right\}
	= \sum_{j} 
	\frac{A_{ij}^{(1)}+A_{ij}^{(2)}}{k_j} \left\{
	\rho_j g(\rho_i)
	-\frac{k_j}{k_i}\rho_i g(\rho_j)
	\right\}\,,
	\label{eq:ODE}
	\end{equation}
	where the second equality has been obtained by exploiting the fact that the multigraph is undirected. Introducing the random walk Laplacian matrix $L$ with elements
	\begin{equation}
	L_{ij} = \frac{A_{ij}^{(1)}+A_{ij}^{(2)}}{k_j} - \delta_{ij}
	\end{equation}  
	allows to rewrite the system of ODEs in a compact form, namely
	\begin{equation}
	\diff{\rho_i}{t} = \sum_{j} L_{ij} \left\{
	\rho_j g(\rho_i) - \frac{k_j}{k_i}\rho_i g(\rho_j)
	\right\}.
	\end{equation}
	By using the property that the eigenvector associated to the $0$ eigenvalue is proportional to the node degree vector, i.e., $(k_1,\dots,k_\Omega)^\top$, the stationary node densities $\rho_i^*$ are found to satisfy~\cite{ACDPFP2018}
	\begin{equation}
	k_i g(\rho_i^*) = c\rho_i^*\, ,
	\end{equation}
	for some constant $c\equiv c(\beta)$ that is uniquely fixed by mass conservation, i.e., $\sum_i \rho_i^* = \beta \Omega$.  
	
	\section{Uncovering the network structure}\label{sec:reconstruction}
	
	Previous work has shown how the above presented nonlinear diffusion process allows to uncover the degree distribution of the corresponding (unknown) graph structure \cite{ACDPFP2018}. We now review and extend their approach by considering, for the sake of pedagogy, the case of multigraphs with two types of links; let us however observe that the method is general enough to deal with more than two attributes for the links. More specifically, we will assume the subgraph associated with the attribute $\mathcal{A}=1$ to be known and will aim at uncovering structural properties of the subgraph associated with the other attribute $\mathcal{A}=2$, under the assumption that the degree distributions of both subgraphs are uncorrelated. The approach relies on an explicit expression of the stationary densities. From now on, we assume $g(x)=1-x$. For such choice, the stationary node densities read
	\begin{equation}
	\rho_i^* = \frac{c(\beta)k_i}{1+c(\beta)k_i}.
	\label{eq:statdens}
	\end{equation}
	The mass conservation then implies
	\begin{equation}
	\sum_{i} \frac{c(\beta)k_i}{1+c(\beta)k_i} = \beta \Omega,
	\end{equation}
	or, equivalently,
	\begin{equation}
	\sum_{k^{(1)},k^{(2)}} p\left(k^{(1)},k^{(2)}\right) \frac{c(\beta) [k^{(1)} + k^{(2)}]}{1+c(\beta)[k^{(1)} + k^{(2)}]} = \beta,
	\end{equation}
	where $p\left(k^{(1)},k^{(2)}\right)$ is the probability that a node will have $k^{(1)}$ links with attribute $\mathcal{A}=1$ and $k^{(2)}$ links with attribute $\mathcal{A}=2$. The degree distributions of the two subgraphs being uncorrelated, the joint probability distribution function factorizes, i.e., $p\big(k^{(1)},k^{(2)}\big) = p_1(k^{(1)})p_2(k^{(2)})$, leading to:
	\begin{equation}
	\sum_{k^{(2)}} \left( \sum_{k^{(1)}} p_1(k^{(1)}) \frac{c(\beta) [k^{(1)} + k^{(2)}]}{1+c(\beta)[k^{(1)} + k^{(2)}]}\right) p_2(k^{(2)})= \beta\, .
	\label{eq:linear}
	\end{equation}
	The latter expression holds for any arbitrary fixed value of $\beta$. Consider thus to perform $M$ experiments (for some integer $M$) each one involving a different number of agents, $\lfloor \beta_i N\Omega \rfloor$,  walking on the multigraph and assume to measure for each experiment the asymptotic average number of walkers sitting on a randomly chosen node (fixed once for all~\footnote{Let us observe that for each value of $\beta_i$, one could in principle choose an arbitrary node to measure $\rho_i^*$, since from Eq.~\eqref{eq:statdens} it is clear that the constant $c(\beta_i)$ can be recovered if we know $\rho_i^*$ and $k_i$. However this would assume a larger knowledge of the network with respect to the case hereby assumed where we only have access to the stationary node density of a given node, yet for distinct values of $\beta$.}) and eventually obtain $c(\beta)$ by using Eq.~\eqref{eq:statdens}. Recalling that the architecture of the network with attribute $\mathcal{A}=1$ is known, we can define the matrix with known entries:
	\begin{equation}
		\label{eq:Fij}
	F_{ij} = \sum_{k^{(1)}} p_1(k^{(1)}) \frac{c(\beta_i) [k^{(1)} + j]}{1+c(\beta_i)[k^{(1)} + j]}\, .
	\end{equation}
	Finally by introducing the vector $\boldsymbol{\beta}:={(\beta_1,\beta_2,\cdots,\beta_M)}^\top$ and writing $\boldsymbol{p_2}:={(p_2(1),p_2(2),\cdots,p_2(M))}^\top$ for the unknown degree distribution of the network with attribute $\mathcal{A}=2$, we can rewrite~\eqref{eq:linear} as 
	\begin{equation}
	F \boldsymbol{p_2} = \boldsymbol{\beta},
	\label{eq:linearsystem}
	\end{equation}
The probability distribution $\boldsymbol{p_2}$ can thus be obtained by solving the linear system~\eqref{eq:linearsystem}. Observe that inverting the matrix $F$ is not recommended, since the matrix $F$ is close to singular. Rather, the solution will be obtained by a least squares method, i.e., by minimizing the $2$-norm $\vert\vert F \boldsymbol{p_2} - \boldsymbol{\beta} \vert \vert_2$. 
	
	Before {proceeding with} the reconstruction of $\boldsymbol{p_2}$, let us for a moment come back to the expression given in Eq.~\eqref{eq:linear}. At low densities, i.e., for $\beta \rightarrow 0$, the stationary node densities, $\rho_i^*$, are close to $0$. It follows, from Eq. (\ref{eq:statdens}), that the constant $c(\beta)$ admits the following expansion $c(\beta)=c_1\beta + c_2\beta^2+c_3\beta^3+O(\beta^4)$ as $\beta \rightarrow 0$. Inserting the latter expression into (\ref{eq:linear}) and gathering terms of the same power in $\beta$, we deduce:
	\begin{equation}
	\begin{split}
	&c_1 = \frac{1}{\langle k_1 \rangle + \langle k_2 \rangle },\\
	&c_2 = \frac{\langle k_1^2 \rangle + \langle k_2^2 \rangle + 2\langle k_1 \rangle \langle k_2 \rangle
	}{
		{\Big( \langle k_1 \rangle + \langle k_2 \rangle \Big)}^3
	},\\
	&c_3 = 2\frac{{\Big( \langle k_1^2 \rangle + \langle k_2^2 \rangle + 2\langle k_1 \rangle \langle k_2 \rangle \Big)}^2
	}{
		{\Big( \langle k_1 \rangle + \langle k_2 \rangle \Big)}^5
	}
	-\frac{ \langle k_1^3 \rangle + \langle k_2^3 \rangle
		+3 \langle k_1^2 \rangle \langle k_2 \rangle 
		+3 \langle k_1 \rangle \langle k_2^2 \rangle
	}{
		{\Big( \langle k_1 \rangle + \langle k_2 \rangle \Big)}^4
	}\, .
	\end{split}
	\label{eq:c123}
	\end{equation}
	The latter expressions provide a direct way of obtaining the degree momenta $\langle k_2^m\rangle = \sum_{k^{(2)}} {(k^{(2)})}^m p_2(k^{(2)})$ of the unknown subgraph, given again the ones of the known network. Indeed, for a given $\beta$, the knowledge of the stationary density of a single (arbitrary) node allows to deduce the corresponding value of the constant $c(\beta)$. By varying $\beta$, we obtain a set of measurement points $(\beta_i,c(\beta_i))$, from which the constants $c_1,c_2,\cdots$ can be deduced by interpolation. The latter constants then allow to obtain the unknown momenta $\langle k_2^m\rangle = \sum_{k^{(2)}} {(k^{(2)})}^m p_2(k^{(2)})$, being the momenta $\langle k_1 ^m \rangle = \sum_{k^{(1)}} {(k^{(1)})}^m p_1(k^{(1)})$ explicitly known. 
	
	{We illustrate the procedure by considering a multigraph made of $100$ nodes, the known subgraph of which is an Erd\H{o}s-Rényi graph, $G_1$, with a probability $q_1=0.15$ of link connection and whose unknown subgraph is also an Erd\H{o}s-Rényi subgraph, this time with a probability $q_2=0.3$ of link connection, of course this information will not be used in the reconstruction strategy. Fig.~\ref{fig:interpolation} (left) shows (in blue) the data points $(\beta_i,c(\beta_i))$ obtained from Eq. (\ref{eq:statdens}) for $0\leq \beta_i \leq \beta_{\text{max}} = 0.2$, along with the cubic polynomial interpolation $c_1\beta + c_2\beta^2 + c_3 \beta^3$ (in dashed red). {From these coefficients, one can obtain the first moment degrees, as shown in the Table provided in Fig. \ref{fig:interpolation}. These momenta are in close agreement with the exact degree momenta provided in the second column (Exact). The following two columns list the degree momenta obtained by solving the linear system \eqref{eq:linearsystem} (Interpolation) and the set of equations~\eqref{eq:c123} (Reconstruction); the values in parenthesis are the relative errors.}
	Let us conclude by observing that the coefficients of the interpolating polynomial depend on the choice of $\beta_{max}$ and so do the reconstructed degree momenta (see Fig.~\ref{fig:interpolation2} in Appendix~\ref{app:deponbeta}).	  
		
\begin{figure}
	\begin{minipage}[c]{0.4\textwidth}
		\centering
		\includegraphics[width=\textwidth]{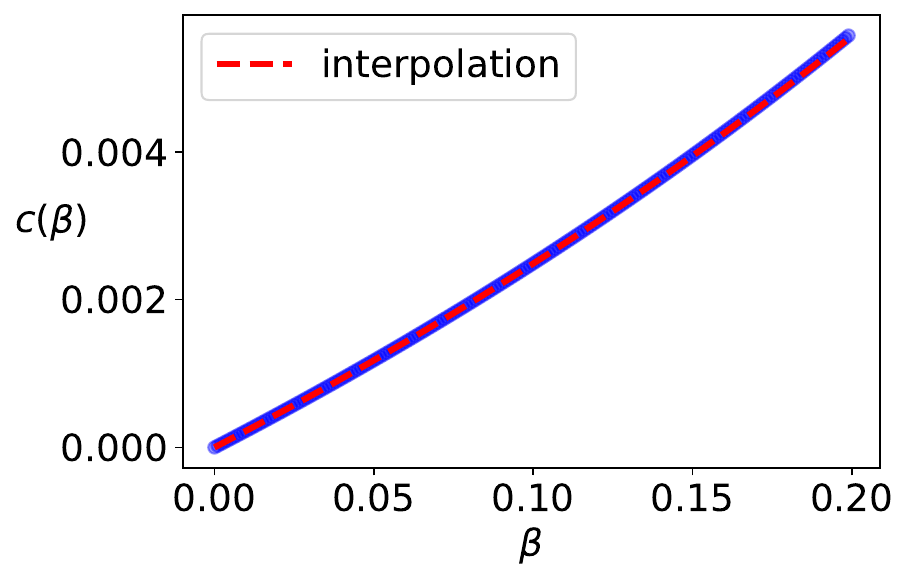}
	\end{minipage}\hspace{0.5cm}
	\begin{minipage}[c]{0.5\textwidth}
		\centering
		\renewcommand{\arraystretch}{1.7}
		\resizebox{\textwidth}{!}{%
			\begin{tabular}{|c|c|c|c|c|}
				\hline
				& Exact & Interpolation & Reconstruction \\
				\hline
				$\langle k_2 \rangle$ & $29.72$ & $29.72$ (0 \%) & $29.72$ (0 \%)\\
				\hline
				$\langle k_2^2 \rangle$ & $907$ & $902$ (0.55 \%) & $904$ (0.22 \%)\\
				\hline
				$\langle k_2^3 \rangle$ & $28\,382$ & $25\,475$ (10.24 \%)& $28\,100$ (1.00 \%)\\
				\hline
			\end{tabular}
		}
	\end{minipage}
	\caption{Reconstruction of the degree momenta of an Erd\H{o}s-Rényi subgraph $G_2$ made of $n=100$ nodes with a probability $q_2=0.3$ of link connection. The known subgraph is an Erd\H{o}s-Rényi graph made of $100$ nodes with probability $q_1=0.15$ of link connection. (Left) Data points $(\beta_i,c(\beta_i))$ along with the cubic polynomial interpolation. { (Right) Comparison between the exact and reconstructed degree momenta $\langle k_2^m\rangle$ of $G_2$ ($m=1,2,3$). The second column gives the exact values of the degree momenta. The third column shows the values obtained from the interpolation procedure, see \eqref{eq:c123} and the fourth column the results obtained by solving the linear system \eqref{eq:linearsystem} to obtain the degree distribution $\bm{p_2}$ and the degree momenta. The relative errors are reported in parenthesis.}}
	\label{fig:interpolation}
\end{figure}

	\section{Impact of the network structure}\label{sec:robustness}
	
	We mentioned in the previous section that the sought degree distribution $p_2(k^{(2)})$ can be found as the solution of a linear problem and that it could be obtained by minimizing the $2$-norm $\vert\vert F \boldsymbol{p_2}-\boldsymbol{\beta}\vert \vert_2$. The purpose of this section is to show the effectiveness of the proposed strategy and to investigate how the accuracy of the reconstruction procedure is influenced by the topologies of the known and unknown subgraphs. To this aim, we will consider three distinct subgraph topologies, namely, Erd\H{o}s-Rényi, scale-free and {Watts-Strogatz}. {We further report in Appendix~\ref{app:bimodal} the results obtained for a bimodal degree distribution.}
	In each case, we will compute the degree distribution by means of a mean-square method. The accuracy of the reconstruction will be assessed by the (relative) error made on the first two momenta of the degree distribution, i.e., $\langle k_2 \rangle $ and $\langle k_2^2\rangle$. This choice is motivated by the fact that the first two momenta of the degree distribution already provide crucial information on the interplay between network structure and dynamics. For instance, the epidemic threshold $\lambda_{c}$ of the SIS model occurs, in the mean-field limit, at $\lambda_{c}=\langle k \rangle / \langle k^2 \rangle$. 
	
	\subsection{Reconstruction of an Erd\H{o}s-Rényi subgraph}
	In this subsection, we assume the unknown subgraph $G_2$ to be given by an Erd\H{o}s-Rényi graph of $\Omega=250$ nodes and probability $q_2$ of connection among nodes, and we are interested in investigating the impact of the known topology of subgraph $G_1$ on the quality of the reconstruction. We first consider the case where $G_1$ is also an Erd\H{o}s-Rényi graph of $\Omega=250$ nodes and probability $q_1$ of links connection. The top panel of the first column of Fig.~\ref{fig:ER} provides the relative error ({all the relative errors will be expressed in percent and} in $\log_{10}$ scale for better visualization) made on the estimated average degree $\langle k_2 \rangle $ of the unknown subgraph $G_2$. Similarly, the bottom panel of the same column provides the relative error for the second moment, $\langle k_2^2 \rangle$. Those errors are computed for a large interval of values of the probabilities $q_1$ and $q_2$ used to build the Erd\H{o}s-R\'enyi graphs.
	{In practice, for each pair of parameters $(q_1,q_2)$, we sampled $10$ multigraphs and computed for each of them the relative error made on the reconstructed degree momenta. The relative errors, obtained for each replica, were then averaged~\footnote{Let us note that one could in principle first average the reconstructed degree momenta and then compute the relative error. However as we are interested in quantifying the quality of a single reconstruction, we decided to first compute the relative error and then average the latter over the set of replicas.}.} In the middle panel, we show, for a specific couple $(q_1,q_2)$ of links connection probabilities, the reconstructed degree distribution (blue symbols) and the comparison with the exact one (red symbols). 
	\begin{figure}[htb!]
		\centering
		\includegraphics[width=\textwidth]{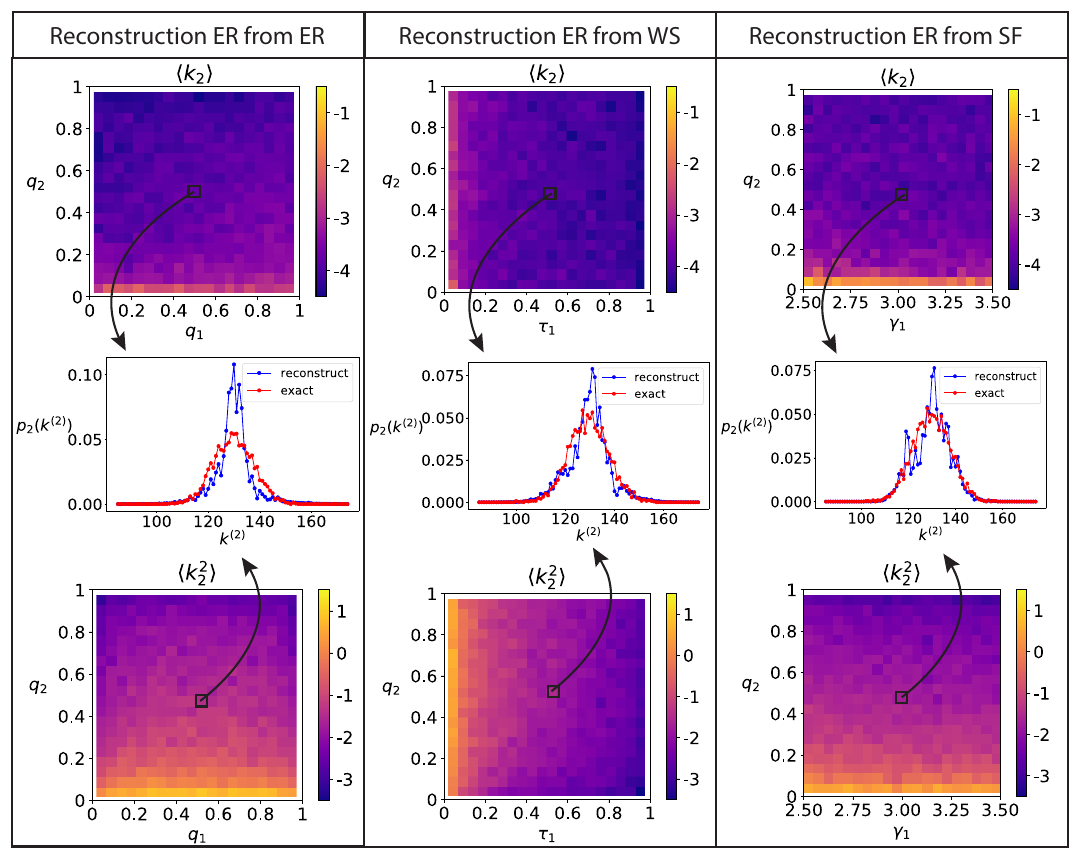}
		\caption{Reconstruction of an Erd\H{o}s-Rényi (ER) subgraph $G_2$ with $\Omega=250$ nodes and probability $q_2$ of links connection. (First column) The known subgraph $G_1$ is an Erd\H{o}s-Rényi graph with $\Omega=250$ nodes and probability $q_1$ of links connection. The relative error (in $\log_{10}$) made on the average degree $\langle k_2 \rangle$ and second moment $\langle k_2^2 \rangle$ are reported on the first and third rows. {For each panel the data were averaged over $10$ independent replicas}. The central panel shows, for some specific values of the parameters $q_1$ and $q_2$ the reconstructed distribution compared with the exact one. (Second column) Reconstruction of $G_2$ by using a {Watts-Strogatz (WS)} graph $G_1$ with probability $\tau_1$ of link rewiring, starting from a ring topology in which each node is connected to its $32$ nearest neighbors. (Third column) Reconstruction of $G_2$ by using a graph $G_1$ with a scale-free (SF) degree distribution with exponent $\gamma_1$, i.e., $p_1(k^{(1)})\sim 1/[k^{(1)}]^{\gamma_1}$.
		}
		\label{fig:ER}
	\end{figure}
	
	We then repeat the same experiment in the case where $G_1$ {is a Wattz-Strogatz} graph (second column) and eventually in the case it has a scale-free degree distribution with exponent $\gamma_1$ (third column). We here briefly comment on the procedure used to generate these subgraph{s}.
	{The Watts-Strogatz graph is obtained by starting with a regular ring lattice of $N$ nodes, with each node connected to its $k$ nearest neighbors. With probability $\tau_1$, each edge $(u,v)$ of the initially regular graph is replaced by a new edge $(u,w)$ where the node $w$ is chosen at random among all the nodes (distinct) from $v$ \cite{WS1998}. In all the analyses shown here, the parameters $N=250$ and $k=32$ are fixed while we vary the parameter $\tau_1$ to investigate the impact of the link rewiring on the accuracy of the reconstruction scheme.} 
	The scale-free graphs were generated according to the Simon model~\cite{bornholdt2001world} which allows to build connected graphs with a power-law degree distribution $p(k)\sim 1/k^{\gamma_1}$ with $\gamma_1 \geq 2$ .
	
	The numerical results reported in Fig.~\ref{fig:ER} suggest that the reconstruction of the first moment is better than for the second one and the goodness of the reconstruction increases with $q_2$, i.e., the larger the connectivity of the unknown graph, the better are the results. {When $G_1$ is a Watts-Strogatz graph (second column), the reconstruction works better when the randomness of the network is increased, i.e., when the links rewiring parameter $\tau_1$ is increased}. In all the three cases, the results suggest that the impact of the known graph $G_1$, as measured by $q_1$, $\mu_1$ or $\gamma_1$, are quite limited. 
	The middle panels confirm such claim, indeed the reconstructed probability distribution reproduces the main feature of the exact distribution, in particular it is centered around the right value, i.e., the average degree, but the widths are a bit narrower than the true ones.

	\subsection{Reconstruction of a scale-free graph}
	We now turn our attention to the reconstruction of a scale-free graph with $\Omega=250$ nodes and exponent $\gamma_2$. Results are reported in Fig. \ref{fig:SF}. The reconstruction of a scale-free graph from the knowledge of an Erd\H{o}s-Rényi graph {or a Watts-Strogatz graph} is quite independent of the involved parameters $q_1$, $\gamma_2$ {and $\tau_1$}, as measured by the relative errors made on $\langle k_2 \rangle$ and $\langle k_2^2 \rangle$ (see first {and second} column{s}). 
	The reconstruction of a scale-free graph from a scale-free graph (third column) is best achieved for large values of $\gamma_1$ and low values of $\gamma_2$. {Overall, the results suggest that the reconstruction works better when the known graph $G_1$ is an Erd\H{o}s-Rényi graph or a Watts-Strogatz graph.}
	\begin{figure}[htb!]
		\centering
		\includegraphics[width=\textwidth]{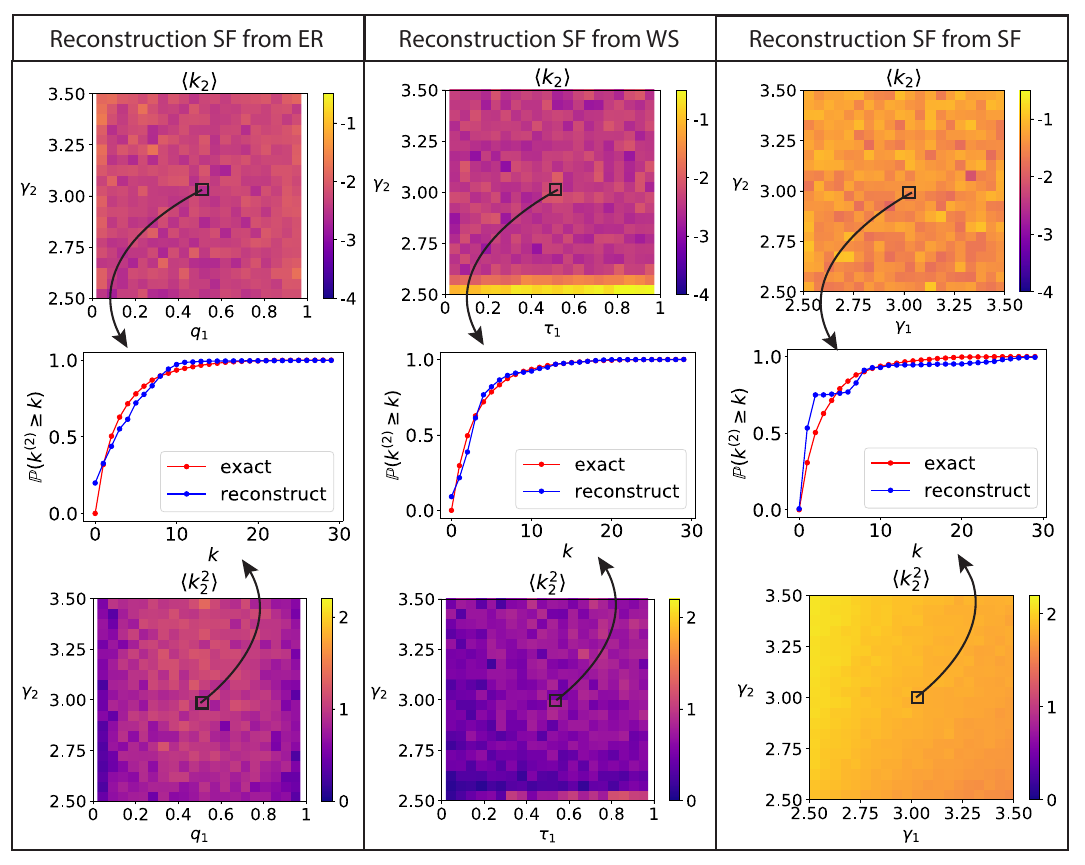}
		\caption{Reconstruction of a subgraph $G_2$ with scale-free degree distribution with exponent $\gamma_2$ and $\Omega=250$ nodes. (First column) The known subgraph $G_1$ is an Erd\H{o}s-Rényi graph with $\Omega=250$ nodes and probability $q_1$ of link connection. The relative error (in $\log_{10}$) made on the average degree $\langle k_2 \rangle$ and second moment $\langle k_2^2 \rangle$ are reported on the first and third rows (averaged over 10 independent configurations). The central panel shows the reconstructed distribution, for specific values of the parameters $\gamma_2$ and $q_1$, averaged over several configurations, and the exact one. (Second column) Same with $G_1$ a Watts-Strogatz graph. (Third column) Same with $G_1$ a scale-free network with exponent $\gamma_1$, i.e., $p_1(k^{(1)})\sim 1/[k^{(1)}]^{\gamma_1}$. 
		}
		\label{fig:SF}
	\end{figure}

\subsection{Reconstruction of a Watts-Strogatz graph}
	We finally consider the reconstruction of a Watts-Strogatz graph $G_2$. As mentioned previously, the graph is obtained by starting with a regular ring lattice of $N=250$ nodes, in which each node is connected to its $32$ nearest neighbors and by rewiring the edges with some probability $\tau_2$. As in the previous sections, we consider the known subgraph $G_1$ to be chosen among an Erd\H{o}s-Rényi graph, a Watts-Strogatz graph and a scale-free graph. The accuracy of the reconstruction is relatively independent from the known graph, see Fig. \ref{fig:WS}. When $G_1$ is a Watts-Strogatz graph (second column), the accuracy decreases for low values of the links rewiring parameters, i.e., when the underlying subgraphs are almost regular. 
	
	\begin{figure}[htb!]
		\centering
		\includegraphics[width=\textwidth]{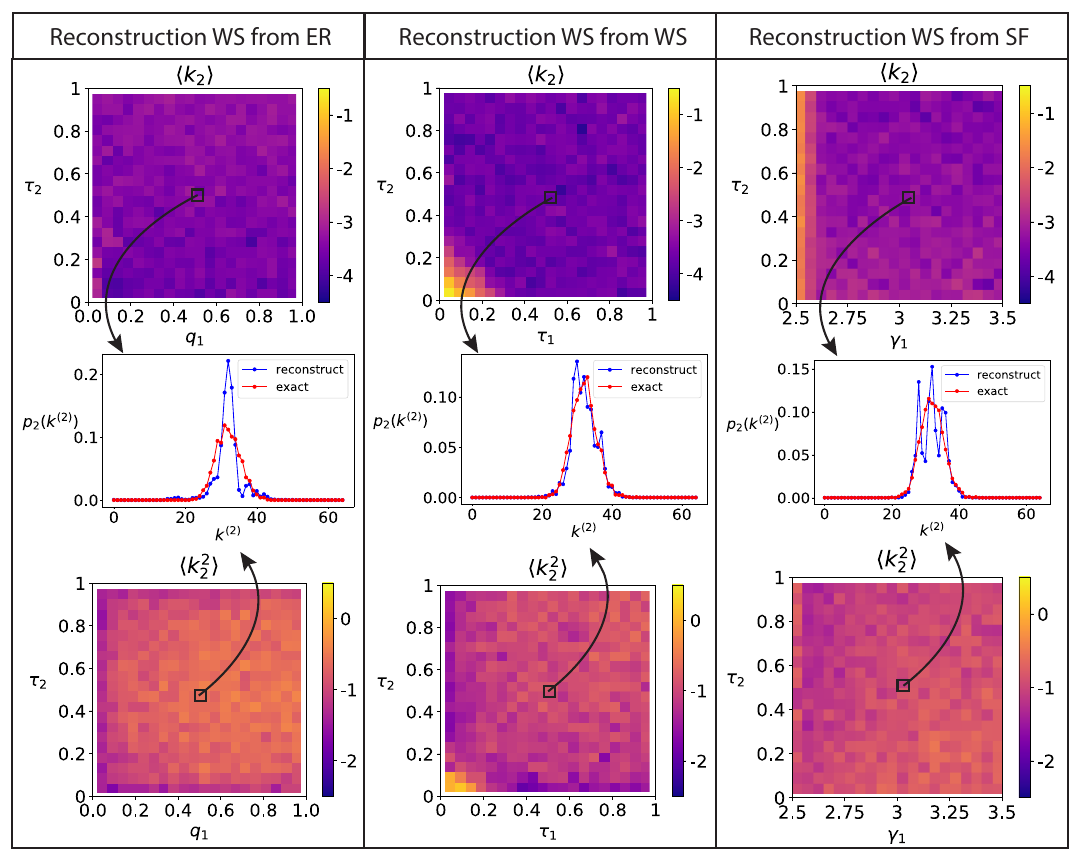}
		\caption{Reconstruction of a Watts-Strogatz graph $G_2$ with $\Omega=250$ nodes, average connectivity $k=32$ and probability $\tau_2$ of links rewiring. (First column) The known subgraph $G_1$ is an Erd\H{o}s-Rényi graph with $\Omega=250$ nodes and probability $q_1$ of link connection. The relative error (in $\log_{10}$) made on the average degree $\langle k_2 \rangle$ and second moment $\langle k_2^2 \rangle$ are reported on the first and third rows (averaged over $10$ independent configurations). The central panel shows, for specific values of the parameters $\tau_2$ and $q_1$, the reconstructed distribution, averaged over several configurations, and compared with the exact one. (Second column) Same with $G_1$ a Watts-Strogatz graph, with average connectivity $k=32$ and probability $\tau_1$ of links rewiring. (Third column) Same with $G_1$ a scale-free network with exponent $\gamma_1$, i.e., $p_1(k^{(1)})\sim 1/[k^{(1)}]^{\gamma_1}$.}
		\label{fig:WS}
	\end{figure}

\subsection{Application to the reconstruction of a real multigraph}

In this subsection, we test the reconstruction algorithm on a real multigraph describing the relationships between the members of a corporate law partnership \cite{Lazega2001,snijders2006new}. Two additional examples are provided in App. \ref{app:real}. The dataset used in the following contains $71$ nodes (i.e., members) and $2223$ links. There are three kinds of links \cite{Lazega2001}:
\begin{itemize}
	\item \textit{Co-work} links between two members indicate that they worked together.
	\item \textit{Advice} links connect two members if one of the two consulted the other for some professional advice.
	\item \textit{Friendship} links between two members indicate that the two of them have socialized outside work.
\end{itemize}
Links in the multigraph carry an orientation. We here discard this information and assume the links to be undirected. As the reconstruction algorithm introduced previously considers two subgraphs, i.e., the known subgraph $G_1$ and the unknown one $G_2$, we grouped the co-work and advice links (which can be viewed as \textit{professional} links). We then proceed with the reconstruction of the degree distribution of the professional links, assuming the subgraph of friendship relationships to be known. The reconstructed degree distribution is shown in Fig. \ref{fig:partnership} (Left) and is found to be in close agreement with the exact one. We then reverse the roles and apply the algorithm to the reconstruction of the degree distribution of the friendship links, under the knowledge of the subgraph of the professional links, see Fig. \ref{fig:partnership} (Right). The reconstruction is less accurate, due to the low density of the subgraph we aimed at reconstructing. In particular, a non-negligible fraction of the nodes have no friendship connections (such nodes have degree $k^{(2)}=0$, but the network is still connected thanks to the links of the other kind). The algorithm is able to predict the presence of such nodes although the estimate ($\approx 0.3$) overestimates the exact value ($\approx 0.2$). We further investigate in App. \ref{app:low_dens} the impact of the density of the unknown subgraph on the accuracy of the reconstruction. Observe that the method hereby used relies on the assumption of absence of correlations among the degree distributions on each subgraph; in App.~\ref{app:correlations} we studied the impact of correlations by using a synthetic network and we observed a strong impact especially in the case the degree distributions are anti-correlated, i.e., the degree distribution is disassortative. {Let us note that the degree momenta of the unknown subgraph can also be obtained by measuring the constant $c(\beta)$ for small $\beta$, and by using Eq. \eqref{eq:c123}. For the reconstruction of the degree distribution of the advice and co-work links, the procedure yields\footnote{We used an interpolation of degree $5$, i.e., $c(\beta) = \sum_{j=1}^5 c_j \beta ^j$, with the constants $c_j$ estimated from the measurement points $(\beta_i,c(\beta_i))$ with $\beta_i$ sampled on the interval $[0,0.02]$ by increment of $0.001$.} the estimate $\langle k_2 \rangle \approx 25.8$ and $\langle k_2^2 \rangle \approx 733$, to be compared with the exact values $\langle k_2 \rangle = 25.6$ and $\langle k_2^2 \rangle = 728 $. For the reconstruction of the degree distribution of the friendship links, we obtain $\langle k_2 \rangle \approx 2.82$ and $\langle k_2^2 \rangle \approx  20.5$, while the exact values are given by $\langle k_2 \rangle = 2.82$ and $\langle k_2^2\rangle  = 15.3 $.}

\begin{figure}[htb!]
	\centering
	\includegraphics[width=0.49\textwidth]{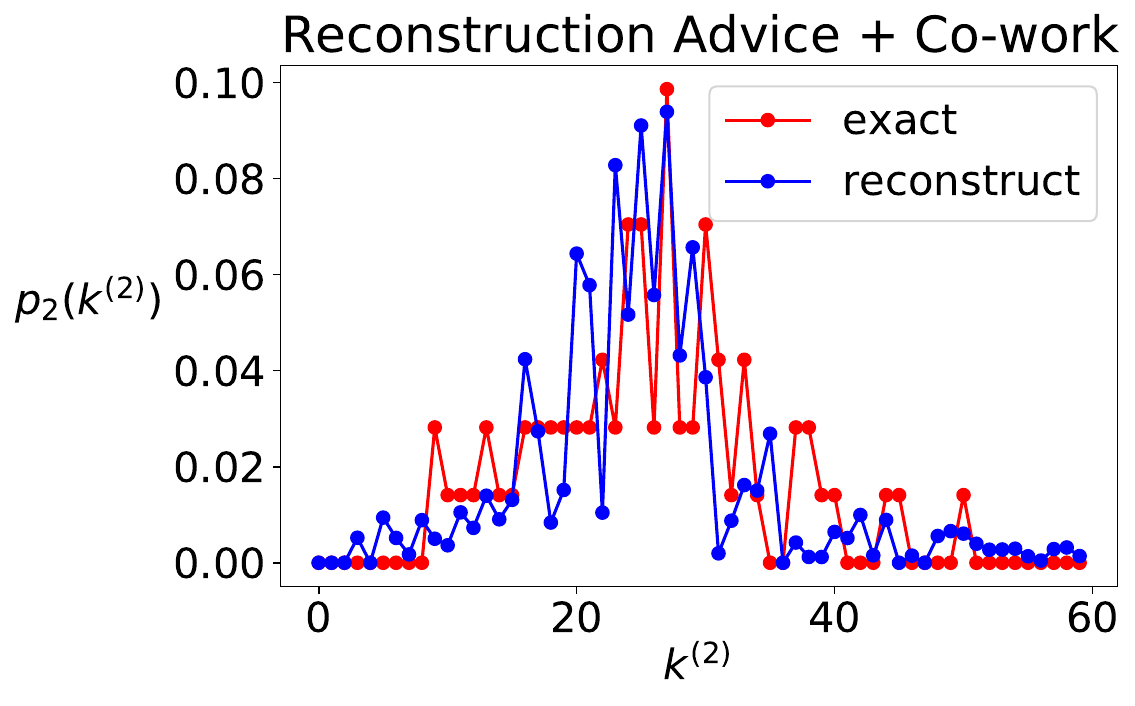}
	\includegraphics[width=0.47\textwidth]{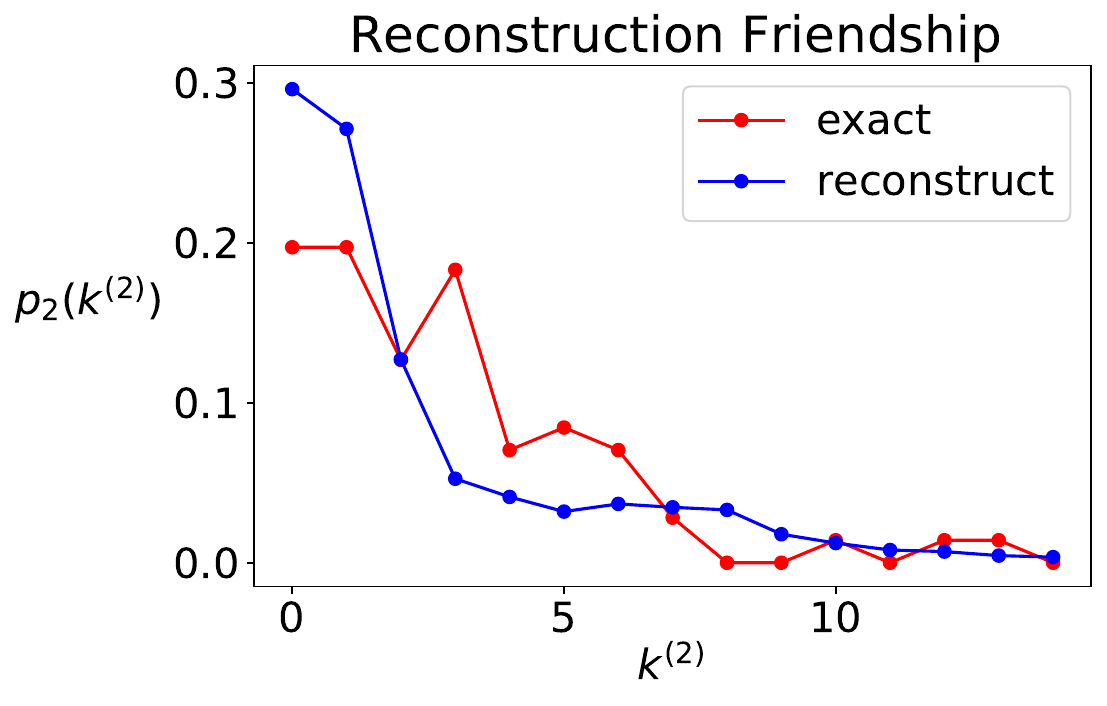}
	\caption{Application to a real corporate law partnership multigraph. Reconstruction of the degree distribution of professional links (co-work and advice links) from the knowledge of the friendship links (Left) and the other way around (Right).}
	\label{fig:partnership}
\end{figure}

	\section{Conclusion}\label{sec:conclusion}
	We here studied a non-linear diffusion process on top of multigraphs, where pairs of nodes can be connected by different types of links. The process was derived from a microscopic description modelling the dispersal of a set of agents under limiting node carrying capacities. As for the case of simple graphs, the stationary node densities show a nonlinear dependence on the node degrees.
	From the explicit knowledge of the stationary node densities, we extended an existing algorithm to reconstruct the degree distribution associated to a given type of links, by assuming the pattern of the other types of interactions to be known.  We first applied the algorithm to the case of synthetic multigraphs, assessing in each case the robustness of the reconstruction by measuring the relative errors on the first two degree momenta. We then applied the algorithm to three real multigraphs; this allowed us to benchmark the method on networks where the assumption of uncorrelated nodes degree is no longer true and thus to study the impact of assortativity on the results. 
	We also considered a second method based on a diluted nonlinear random walk,  i.e., once a smaller and smaller number of walkers is allowed to travel across the multigraph links, in this way we have been able to directly measure the first and second degree momenta without the need to compute the degree distribution. This algorithm has been successfully tested both on synthetic than empirical networks.
	Future work could include an extension of the reconstruction algorithm to the case of multilayer networks in which there are both interlayer and intralayer links. 
	
	\section{Acknowledgements}
	
	This work was supported by the Fonds de la Recherche Scientifique—FNRS (Grant FC 38477 to J.-F. d K). Part of the results were obtained using the computational resources provided by the ‘‘Consortium des Equipements de Calcul Intensif’’ (CECI), funded by the Fonds de la Recherche Scientifique de Belgique (F.R.S.-FNRS) under Grant No. 2.5020.11 and by the Walloon Region.
	
	\clearpage

	\bibliographystyle{plain}

	\clearpage
	\appendix
	\section{Derivation of the ODE from the M-equation}
	\label{app:ME2ODE}
	The purpose of this appendix is to derive the system of ordinary differential equations given in Eq. \eqref{eq:ODE} and reproduced below:
	\begin{equation}
	\diff{\rho_i}{t} = \sum_{j} 
	\left\{
	\rho_j\frac{A_{ji}^{(1)}+A_{ji}^{(2)}}{k_j}g(\rho_i)
	-\rho_i\frac{A_{ij}^{(1)}+A_{ij}^{(2)}}{k_i}g(\rho_j)
	\right\}
	= \sum_{j} 
	\frac{A_{ij}^{(1)}+A_{ij}^{(2)}}{k_j} \left\{
	\rho_j g(\rho_i)
	-\frac{k_j}{k_i}\rho_i g(\rho_j)
	\right\}\,,
	\end{equation}
	from the M-\textit{equation}
	\begin{equation}
	\diff{\probP{}(\mathbf{n},t)}{t} 
	=
	\sum_{\mathbf{n'} \neq \mathbf{n}} 
	\left \{
	T(\mathbf{n} \vert \mathbf{n'})\probP{}(\mathbf{n'},t)
	-T(\mathbf{n'} \vert \mathbf{n})\probP{}(\mathbf{n},t)
	\right\},
	\end{equation}
	with transition rates
	\begin{equation}
	T(n_i-1,n_j+1\vert n_i,n_j) = \frac{A_{ij}^{(1)}+A_{ij}^{(2)}}{k_i} \frac{n_i}{N} g\Big(\frac{n_j}{N}\Big).
	\end{equation}
	We refer the reader to the main text for more details about the notations used. The procedure used is quite standard, see e.g. \cite{CAFL2020}.
	We first recall that the average number of individuals found in node $i$ at time $t$ is given by $\langle n_i(t) \rangle = \sum_{\mathbf{n}} n_i(t) \probP{}(\mathbf{n},t)$. 
	As during an infinitesimal time interval, at most one random walker will have moved, the states $\mathbf{n}$ and $\mathbf{n'}$ will have identical components, except for two of them, corresponding to the nodes involved in the transition. More precisely, if the state before transition has components $\mathbf{n} = (n_1,\cdots,n_i,\cdots,n_j,\cdots,n_\Omega)$, then after the transition, the new state will have components $\mathbf{n'} = (n_1,\cdots,n_i\pm 1,\cdots,n_j\mp1,\cdots,n_\Omega)$. We thus obtain:
	\begin{equation}
	\begin{split}
		\diff{\langle n_i(t) \rangle}{t}  = &\sum_{n_i}\sum_{j, n_j}n_i \Big[
		-T(n_i-1,n_j+1\vert n_i,n_j) \probP{}(n_i,n_j,t)
		+T(n_i,n_j\vert n_i+1,n_j-1) \probP{}(n_i+1,n_j-1,t)
		\Big]\\
		+&\sum_{n_i}\sum_{j, n_j}n_i \Big[
		-T(n_i+1,n_j-1\vert n_i,n_j) \probP{}(n_i,n_j,t)
		+T(n_i,n_j\vert n_i-1,n_j+1) \probP{}(n_i-1,n_j+1,t)
		\Big].
	\end{split}
	\label{eq:2sum}
	\end{equation}
	In the above expression, $T(n_i-1,n_j+1\vert n_i,n_j)\probP{}(n_i,n_j,t)$ denotes the transition rate associated to the jump of an agent from node $i$, containing initially $n_i$ agents, to node $j$, containing initially $n_j$ agents. The three other terms carry a similar interpretation. Playing with the summation indices in Eq. \eqref{eq:2sum} leads to the following expression:
	\begin{equation}
	\begin{split}
		\diff{\langle n_i(t) \rangle}{t} &= \sum_{j}\Big(
		\langle T(n_i+1,n_j-1\vert n_i,n_j) \rangle
		-\langle T(n_i-1,n_j+1\vert n_i,n_j) \rangle
		\Big)\\
		&= \sum_{j}\Big(
		\frac{A_{ji}^{(1)}+A_{ji}^{(2)}}{k_j} \langle\frac{n_j}{N} g(\frac{n_i}{N}) \rangle
		-\frac{A_{ji}^{(1)}+A_{ji}^{(2)}}{k_i} \langle\frac{n_i}{N} g(\frac{n_j}{N}) \rangle
		\Big)
	\end{split}
	\end{equation} 
	Rescaling time $t\rightarrow t N$ and taking the limit $N \rightarrow + \infty$ allows to neglect the correlations and leads to:
	\begin{equation}
	\diff{\rho_i}{t} = \sum_{j} 
	\left\{
	\rho_j\frac{A_{ji}^{(1)}+A_{ji}^{(2)}}{k_j}g(\rho_i)
	-\rho_i\frac{A_{ij}^{(1)}+A_{ij}^{(2)}}{k_i}g(\rho_j)
	\right\},
	\end{equation}
	as stated in Eq. \eqref{eq:ODE}.
	
	\section{Dependence of the degree momenta on the mass parameter $\beta$}
	\label{app:deponbeta}
	
	In this appendix, we comment on the impact that the mass parameter $\beta$ has on the reconstruction of the degree momenta, see Eq. (\ref{eq:c123}). We recall that the latter are obtained by first computing the constant $c(\beta)$ for distinct values of $\beta$ in the interval $[0,\beta_{\text{max}}]$ and then by extrapolating $c(\beta)$ by a polynomial whose coefficients allow to obtain the desired degree momenta. In Fig. \ref{fig:interpolation2}, we compare, as a function of $\beta_{\text {max}}$, the first and second degree momenta, as reconstructed from Eq. (\ref{eq:c123}), with their exact values. The agreement is very satisfactory except for very low values of $\beta_{\text{max}}$.}
	  
	\begin{figure}[htb!]
		\centering
		\includegraphics[width=0.34\textwidth]{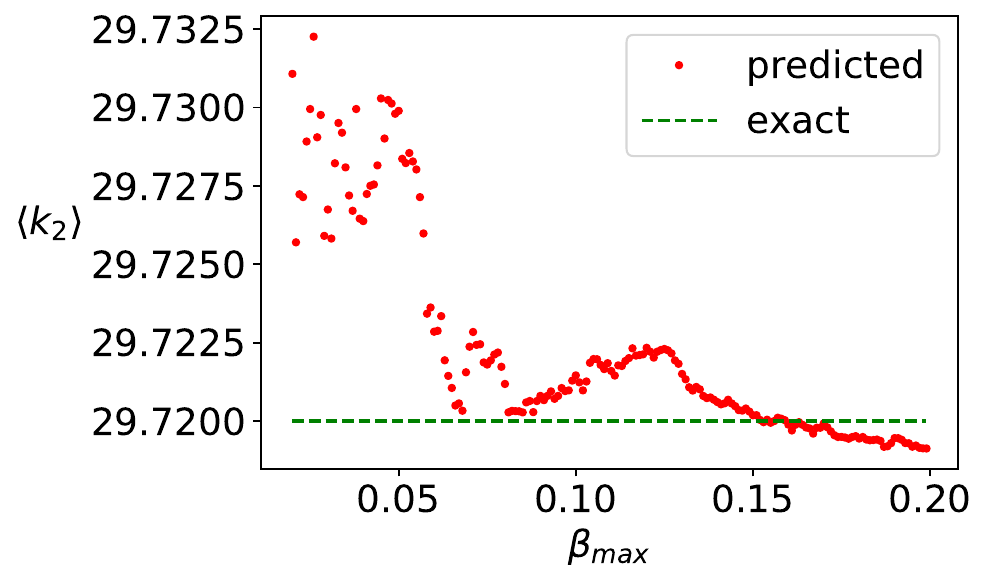}
		\includegraphics[width=0.32\textwidth]{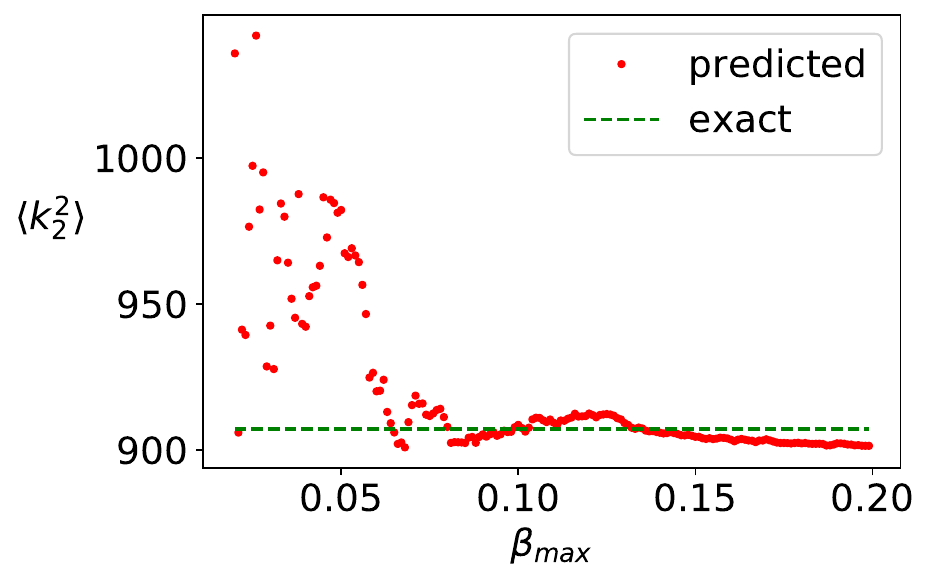}
		\caption{Reconstruction of the degree momenta of an Erd\H{o}s-Rényi subgraph $G_2$ made of $n=100$ nodes with a probability $q_2=0.3$ of link connection. The known subgraph is an Erd\H{o}s-Rényi graph made of $100$ nodes with probability $q_1=0.15$ of link connection. (Left panel) Comparison between the predicted and exact first moment $\langle k_2 \rangle$, as obtained from the interpolation on the interval $0\leq \beta_i \leq \beta_{\text{max}}$. (Right panel) Same for the second moment $\langle k_2^2 \rangle$.}
		\label{fig:interpolation2}
	\end{figure}

	\section{Reconstruction in the case of bimodal degree distribution}\label{app:bimodal}
	In this appendix, we investigate the accuracy of the reconstruction procedure described in the main text when the subgraph $G_1$ and/or the subgraph $G_2$ has (have) a bimodal degree distribution.  
	To construct a subgraph $G_1$ with bimodal degree distribution, we proceed as follows. We first generate the sequence of nodes' degrees, with half of the nodes' degrees sampled from a Gaussian distribution~\footnote{We here assume nodes' degrees to be positive integers between $1$ and $\Omega-1$, with $\Omega$ the number of nodes. The numbers randomly sampled from the Gaussian distribution are thus rounded to their nearest integer value and values outside the interval $(1,\Omega)$, if any, are rejected during the sampling procedure.} with mean $\mu=35$ and standard deviation $\sigma = 10$ , while the remaining nodes degrees are sampled from a Gaussian distribution centered at $\mu_1$ with standard deviation $\sigma = 10$, being $\mu_1$ a varying parameter. Once the degree sequence has been generated, the subgraph $G_1$ is built by using the configuration model with the previously generated sequence of nodes' degrees. We delete self-loops and multiple edges so that the adjacency matrix $A^{(1)}$ associated to the known subgraph has binary entries~\footnote{As a consequence of the supression of self loops and multiple edges, the degree connectivity of the resulting graph is slightly reduced and, in particular, the two peaks of the observed degree distribution $p_2(k^{(2)})$ will be slightly shifted towards lower values of $^{(2)}$. Let us however note that one could still use the reconstruction algorithm presented in the main text in presence of multiple edges.}.
	
\begin{figure}[htb!]
		\centering
		\includegraphics[width=\textwidth]{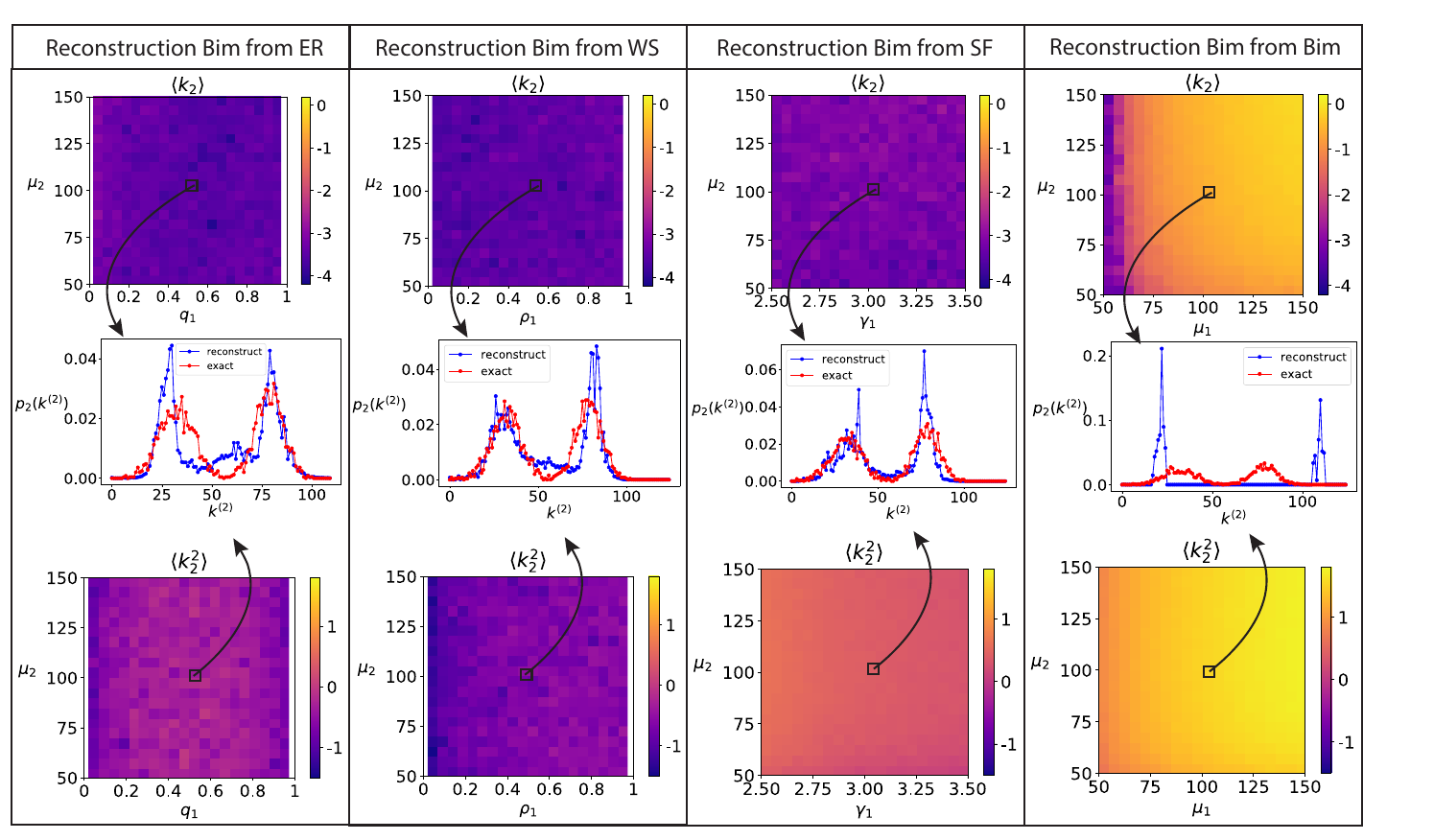}
		\caption{Reconstruction of a subgraph $G_2$ with bimodal degree distribution (Bim) centred at $35$ and $\mu_2$ with variance $10$ and $\Omega=250$ nodes. (First column) The known subgraph $G_1$ is an Erd\H{o}s-Rényi graph with $\Omega=250$ nodes and probability $q_1$ of link connection. The relative error (in $\log_{10}$) made on the average degree $\langle k_2 \rangle$ and second moment $\langle k_2^2 \rangle$ are reported on the first and third rows (averaged over 10 independent configurations). The central panel shows, for specific values of the parameters $\mu_2$ and $q_1$, the reconstructed distribution, averaged over several configurations, and compared with the exact one. (Second column) Same with $G_1$ a Watts-Strogatz graph with probability $\tau_1$ of links rewiring. (Third column) Same with $G_1$ a scale-free network with exponent $\gamma_1$, i.e., $p_1(k^{(1)})\sim 1/[k^{(1)}]^{\gamma_1}$. (Fourth column) Same with $G_1$ having a bimodal degree distribution centred at $35$ and $\mu_1$, with variance $10$.}
		\label{fig:Bimodal}
	\end{figure}

	In Fig. \ref{fig:Bimodal}, we show the reconstruction of a graph $G_2$ with bimodal degree distribution, for four distinct topologies of the known subgraph $G_1$, namely, Erd\H{o}s-Rényi (ER), Watts-Strogatz (WS), scale-free (SF) and bimodal (Bim). The reconstruction strategy works better when the known subgraph is an Erd\H{o}s-Rényi, Watts-Strogatz or a scale-free graph. In contrast, the exact and reconstructed degree distributions deviate significantly when $G_1$ has a bimodal degree distribution (see fourth column) although the bimodality is still predicted.
	
\begin{figure}[htb!]
		\centering
		\includegraphics[width=\textwidth]{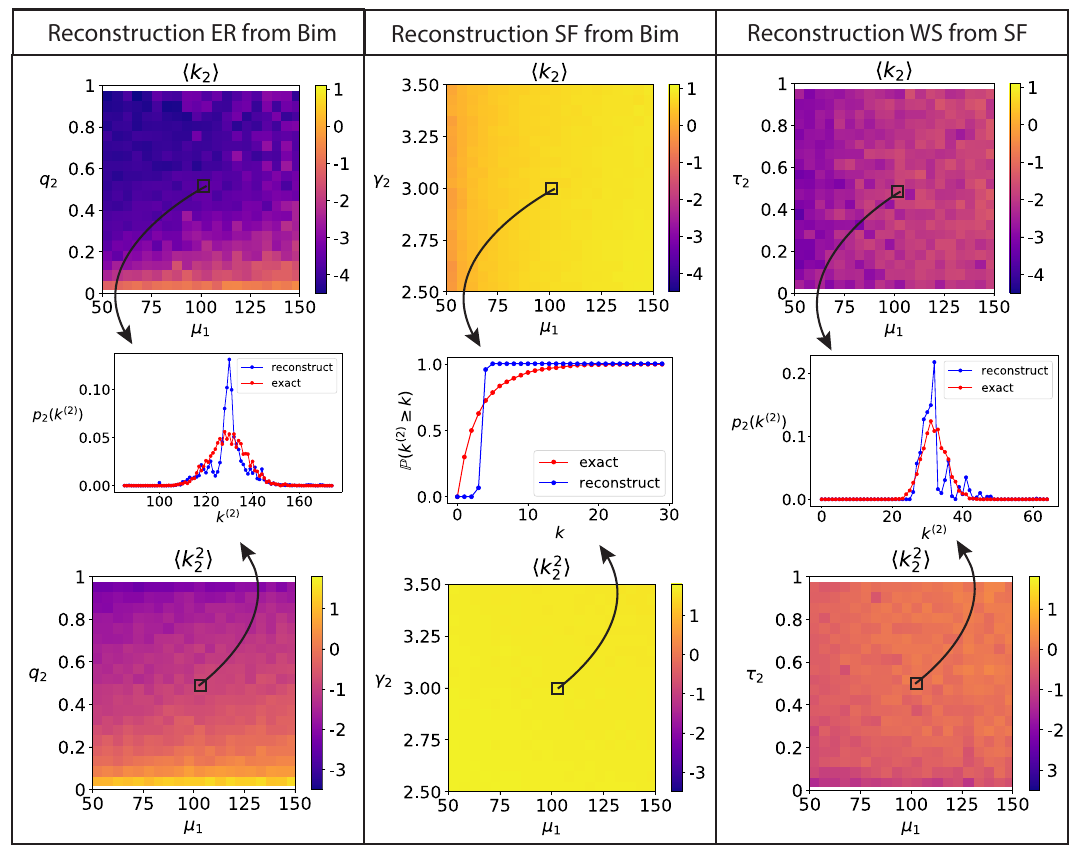}
		\caption{Reconstruction of a subgraph $G_2$ when the known subgraph $G_1$ has bimodal degree distribution (Bim) centred at $35$ and $\mu_2$ with variance $10$ and $\Omega=250$ nodes. (First column) The unknown subgraph $G_2$ is an Erd\H{o}s-Rényi graph with $\Omega=250$ nodes and probability $q_2$ of link connection. The relative error (in $\log_{10}$) made on the average degree $\langle k_2 \rangle$ and second moment $\langle k_2^2 \rangle$ are reported on the first and third rows (averaged over 10 independent configurations). The central panel shows, for specific values of the parameters $\mu_1$ and $q_2$, the reconstructed distribution, averaged over several configurations, and compared with the exact one. (Second column) Same with $G_2$ a scale-free network with exponent $\gamma_2$, i.e., $p_2(k^{(2)})\sim 1/[k^{(2)}]^{\gamma_1}$. (Third column) Watts-Strogatz graph with probability $\tau_2$ of links rewiring.}
		\label{fig:knownBimodal}
	\end{figure}

	Fig. \ref{fig:knownBimodal} shows the accuracy of the reconstruction of a graph when the known subgraph has bimodal degree distribution. The reconstruction works better if the unknown subgraph is an Erd\H{o}s-Rényi (first column) or a Watts-Strogatz (third column).
	
	\clearpage
	\section{Reconstruction of low density networks}\label{app:low_dens}
\begin{figure}[htb!]
		\centering
		\includegraphics[width=\textwidth]{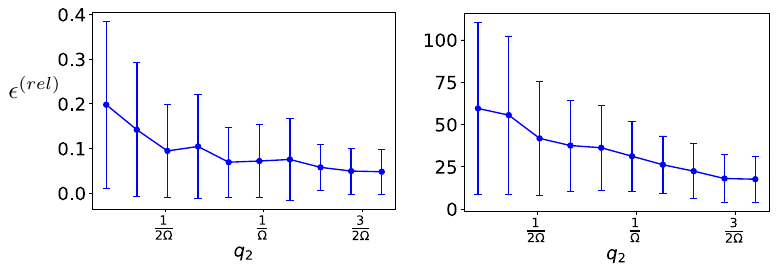}
		\caption{Reconstruction of an Erd\H{o}s-Rényi subgraph $G_2$ of $\Omega=100$ nodes with link connection $q_2$ from the knowledge of an Erd\H{o}s-Rényi subgraph $G_2$ with link connection $q_1=0.5$. In the left panel we show the relative error $\varepsilon^{(rel)}$ on $\langle k_2\rangle$ while in the right panel the one for $\langle k^2_2\rangle$. The data were averaged over $100$ independent replicas. Vertical bars represent the standard deviations.}
		\label{fig:sparseER}
	\end{figure}

	In this appendix, we investigate the accuracy of the reconstruction of low densely connected subgraphs. To this goal, we construct synthetic multigraphs with $G_1$ (known subgraph) and $G_2$ (unknown subgraph) two Erd\H{o}s-Rényi graphs with probabilities $q_1=1\gg 1/\Omega$ and $q_2\sim 1/\Omega$ of link connections. In this way the known subgraph is densely connected while the unknown subgraph is sparse, i.e., the node degrees $k^{(2)}$ are low and possibly zero. In Fig. \ref{fig:sparseER}, we show the relative errors on the average reconstructed degree momenta $\langle k_2 \rangle$ and $\langle k_2^2 \rangle$ as a function of $q_2$ with $\Omega=100$ nodes. Vertical bars indicate the standard deviations deduced from $100$ experiments. As can be seen from this Figure, the averaged relative error and the standard deviations increase as the parameter $q_2$ is reduced, implying thus some limitation of the reconstruction method in presence of zero degree nodes and low density of connections.
	
	\section{Reconstruction of real multigraphs}\label{app:real}
\begin{figure}[htb!]
	\centering
	\includegraphics[width=0.8\textwidth]{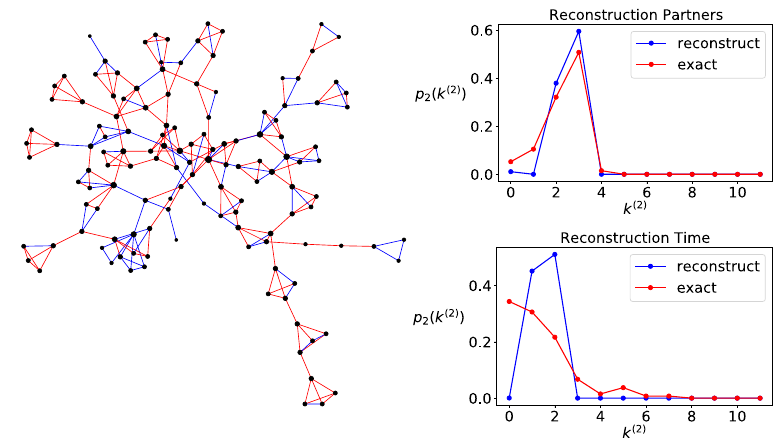}
	\caption{(Left) Cooperation multigraph. Blue links correspond to \textit{Time} links while red links identify \textit{Partners} links. (Right) The top panel (resp. bottom) gives the reconstruction of the degree distribution corresponding to the \textit{Partners} (resp. \textit{Time}) links from the knowledge of the degree distribution of the \textit{Time} (resp. \textit{Partners}) links.}
	\label{fig:3_Student}
\end{figure}

	We further investigate in this Appendix the reconstruction of real multigraphs.
	The first instance is a cooperation multigraph among students in a course \cite{fire2012predicting}. Nodes correspond to students and links identify cooperation between them. Three types of cooperation links were identified, whose attributes are denoted by \textit{Time}, \textit{Computer} and \textit{Partners}. Partners links connect pairs of students that submitted their assignments together, computer links connect pairs of students that used the same computer for their (online) assignments and Time links connect pairs of students that had accessed the course's website almost at the same time. We refer the reader to \cite{fire2012predicting} for more details about the procedure used to collect the data and construct the multigraph. As there are fewer \textit{Computer} links, we discard them and only focus on the Time and Partners links, dealing thus with a multigraph with two attributes.  Moreover, we restrict our attention to the largest connected component\footnote{There must exist a path joining any pair of nodes in the connected component.} of the resulting multigraph, shown in Fig. \ref{fig:3_Student} (Left). The latter contains $134$ nodes and $484$ links, of which $312$ are Partners links and $172$ are Time links. We then proceed with the reconstruction of the degree distribution corresponding to the Partners links, assuming the degree distribution of the \textit{Time} links to be known, see top right panel of Fig. \ref{fig:3_Student}. We then repeat the experiment for the reconstruction of the degree distribution of the Time links, see bottom right panel of Fig. \ref{fig:3_Student}. In the latter case, the reconstruction is less satisfactory. The discrepancy observed might be the consequence of the correlation between the degree sequences of both subgraphs. The Pearson correlation coefficient measuring the correlation between both degree sequences is given by $r\approx -0.5$, which indicates a diassortative behavior: nodes with few Partners links tend to have more Time links and vice versa. We further investigate in Appendix \ref{app:correlations} the impact of degree-degree correlations on the reconstruction accuracy.
{The degree momenta of the unknown subgraph can also be inferred by measuring the constant $c(\beta)$ for small $\beta$, and by using Eq. \eqref{eq:c123}. For the reconstruction of the degree distribution of the Partners links, the procedure yields the estimate $\langle k_2 \rangle \approx 2.33$ and $\langle k_2^2 \rangle \approx 5.17$, to be compared with the exact values $\langle k_2 \rangle = 2.33 $ and $\langle k_2^2 \rangle = 6.19 $. For the reconstruction of the degree distribution of the Time links, we obtain $\langle k_2 \rangle \approx 1.28 $ and $\langle k_2^2 \rangle \approx 2.56 $ while the exact values are given by $\langle k_2 \rangle = 1.28$ and $\langle k_2^2 \rangle = 3.58$.}

\begin{figure}[htb!]
	\centering
	\includegraphics[width=0.8\textwidth]{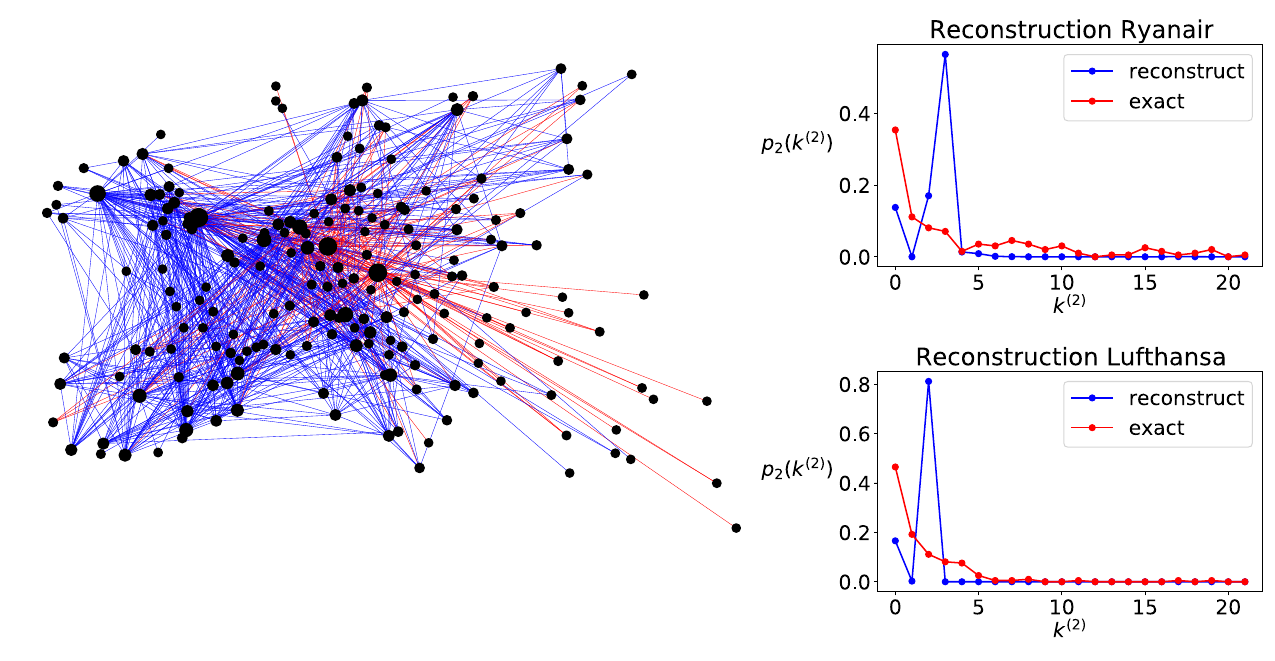}
	\caption{(Left) European air transportation multigraph. Links in blue (resp. red) indicate flight routes operated by Lufthansa (resp. Ryanair) (Right) The top panel (resp. bottom) gives the reconstruction of the degree distribution corresponding to Ryanair's (resp. Lufthansa's) links from the knowledge of the degree distribution of Lufthansa's (resp. Ryanair's) links.}
	\label{fig:3_Airport}
\end{figure}

The second instance is the European air transportation multigraph where links' attributes are commercial arilines \cite{cardillo2013emergence}. We here restrict our attention to two operating companies, namely Lufthansa and Ryanair.  The multigraph is shown in Fig. \ref{fig:3_Airport} (Left). It contains $198$ nodes connected by $1690$ links, of which $488$ are from the Lufthansa operating company and the others from Ryanair. As before, we restrict our attention to the largest connected component. The top right panel of Fig. \ref{fig:3_Airport} shows the reconstructed degree distribution corresponding to Ryanair's links, assuming the degree distribution for Lufthansa's links to be known while the bottom right panel shows the other way around. The discrepancy observed between the exact and reconstructed degree distributions might in part be due to the presence of low-degree nodes and the degree-degree correlations, see Appendices \ref{app:low_dens} and \ref{app:correlations}. 
{Again, the degree momenta of the unknown subgraph can be inferred by measuring the constant $c(\beta)$ for small $\beta$, and by using Eq. \eqref{eq:c123}. For the reconstruction of the degree distribution of Ryanair's links, the procedure yields the estimate $\langle k_2 \rangle \approx 6.07$ and $\langle k_2^2 \rangle \approx 120$, to be compared with the exact values $\langle k_2 \rangle = 6.07 $ and $\langle k_2^2 \rangle = 143 $. For the reconstruction of the degree distribution of the Lufthansa links, we obtain $\langle k_2 \rangle \approx 2.46 $ and $\langle k_2^2 \rangle \approx 54.8 $ while the exact values are given by $\langle k_2 \rangle = 2.46$ and $\langle k_2^2 \rangle = 77.8$.}

	\section{Impact of the degree-degree correlations on the reconstruction}\label{app:correlations}
	The reconstruction procedure described in the main text assumes the degree distributions of the known and unknown subgraphs $G_1$ and $G_2$ to be uncorrelated. This assumption allows to factorize the joint probability degree distribution function $p(k^{(1)},k^{(2)}) = p_1(k^{(1)}) p_2(k^{(2)})$, from which the unknown degree distribution $p_2(k^{(2)})$ is recovered as the solution of a linear system, see Eq.~\eqref{eq:linearsystem}. When the nodes degrees of the known subgraph correlate with that of the unknown subgraph, factorizing the joint probability distribution is no longer exact; it is an approximation that may induce some errors on the reconstruction outcomes. The purpose of this Appendix is precisely to investigate that aspect. To quantify the level of degree-degree correlations, we will use the Pearson correlation coefficient $r$. For the case at hand, the latter coefficient is given by:
	\begin{equation}
		r = \frac{ 
				\langle k_1 k_2 \rangle - \langle k_1 \rangle \langle k_2 \rangle
				}{
				\sigma_1 \sigma_2
				},
	\end{equation}
	where $\langle k_1 k_2 \rangle=\sum_{k^{(1)}, k^{(2)}} k^{(1)} k^{(2)} p(k^{(1)},k^{(2)}) $, $\langle k_1 \rangle=\sum_{k^{(1)}} k^{(1)} p(k^{(1)})$ (and similarly for $\langle k_2 \rangle$) and $\sigma_1, \sigma_2$ are the standard deviations associated to the nodes degrees of $G_1$ and $G_2$.
	
	To investigate the impact of degree-degree correlations on the reconstruction accuracy, we construct correlated multigraphs with a prescribed Pearson correlation coefficient $r$. The algorithm we used is similar to the one described in Section VII. A. in \cite{NL2015}. We start by constructing two synthetic graphs $G_1$ and $G_2$ (from some graph ensembles), each of which has $\Omega$ nodes, labelled $1,2,\cdots,\Omega$. Keeping the graph structure of $G_1$ fixed, the idea is to modify progressively the structure of $G_2$ so as to get closer and closer to the desired Pearson correlation coefficient. In practice, the algorithm will stop when the observed Pearson correlation and the target one differ by a quantity smaller than some tolerance threshold $\epsilon \ll 1$ set by the user. 
	
	We now describe in more details the algorithm. Suppose thus that we want to generate a multigraph with a Pearson correlation coefficient $r$ given above. Starting with two random graphs $G_1$ and $G_2$, we compute the difference $\Delta r$, in absolute value, between the observed Pearson correlation coefficient and the one we would like to obtain. If $\Delta r$ is less than the tolerance threshold $\epsilon$, the algorithm outputs the multigraph. Otherwise, we select a pair of nodes of $G_2$ at random, say $i$ and $j$, and measure the new Pearson correlation coefficient we would obtain by swapping the labels of both nodes, i.e., by permuting rows $i$ and $j$ and columns $i$ and $j$ of the adjacency matrix $A^{(2)}$. If the operation would result in a smaller value $\Delta_r^{\text{new}}$ compared to $\Delta r$, it is accepted with probability $1$. If not, namely, if $\Delta_r^{\text{new}}>\Delta r$ there is still a small probability $e^{-(\Delta_r^{\text{new}} - \Delta r)/\lambda}$ to accept the modification, with $\lambda >0$ some parameter. We repeat the process until $\Delta r$ is smaller than $\epsilon$. Let us note that the graph structure of $G_1$ is preserved throughout the process.
	 
	The above algorithm generates correlated multigraphs. In particular, positive values of $r$ indicate that high-degree nodes in $G_1$ tend to be also high-degree nodes in $G_2$, while, in contrast, negative values of $r$ imply the opposite phenomenon, namely, high degree nodes in $G_1$ have a greater probability to be low degree nodes in $G_2$.
	
\begin{figure}[htb!]
		\centering
		\includegraphics[width=\textwidth]{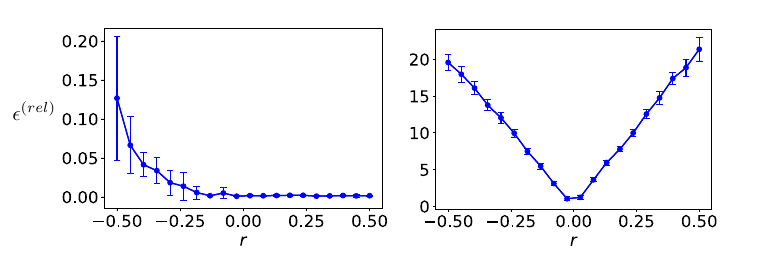}
		\caption{Relative errors $\varepsilon^{(rel)}$ made on $\langle k_2 \rangle$ (left panel) and $\langle k_2^2 \rangle$ (right panel) with $G_1$ (known graph) and $G_2$ (unknown graph) two Barabási-Albert subgraphs of $\Omega=100$ nodes, constructed by preferential attachment: new nodes are added to the graph and connect to $10$ existing nodes. The degree sequences of both subgraphs are correlated, with a Pearson correlation coefficient $r$ (x-axis). Each point was averaged over $10$ independent replicas, with the standard deviations reported by the vertical bars. The parameter $\lambda$ was set to $0.0001$ (see the text for its meaning).}
		\label{fig:corr_BA}
	\end{figure}

\begin{figure}[htb!]
		\centering
		\includegraphics[width=\textwidth]{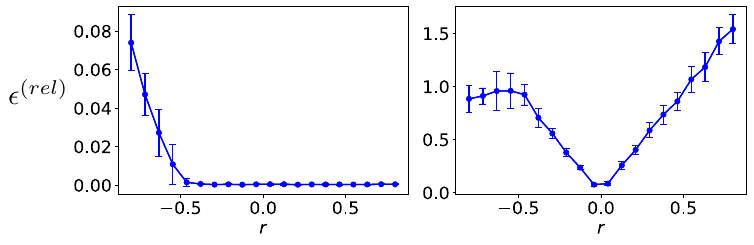}
		\caption{Relative errors $\varepsilon^{(rel)}$ made on $\langle k_2 \rangle$ (left panel) and $\langle k_2^2 \rangle$ (right panel) with $G_1$ (known graph) and $G_2$ (unknown graph) two Erd\H{o}s-Rényi subgraphs of $\Omega=100$ nodes and probability $q_1=1/2$ and $q_2=1/2$ of link connection. The degree sequences of both subgraphs are correlated, with a Pearson correlation coefficient $r$ (x-axis). Each point was averaged over $10$ independent replicas, with the standard deviations reported by the vertical bars. The parameter $\lambda$ was set to $0.0001$ (see the text for its meaning).}
		\label{fig:corr_ER}
	\end{figure}

	We now apply the above algorithm to generate correlated multigraphs and measure the relative errors made on the reconstructed degree momenta $\langle k_2 \rangle$ and $\langle k_2^2 \rangle$. We first assume the graphs $G_1$ and $G_2$ to be Barabási-Albert graphs of $\Omega=100$ nodes. Figure~\ref{fig:corr_BA} gives the relative errors made on the average degree $\langle k_2 \rangle$ (left panel) and second degree moment $\langle k_2^2 \rangle$ (right panel), as a function of the Pearson correlation coefficient $r$. As can be seen from the left panel, the relative error made on $\langle k_2 \rangle$ is almost zero for positive values of the Pearson correlation coefficient. For negative values of $r$, the smaller $r$ the larger the relative error. The right panel shows that the larger $\vert r \vert$ and the larger the relative error. {A similar behavior is observed if the known and unknown subgraphs are Erd\H{o}s-Rényi graphs, see Fig.~\ref{fig:corr_ER}.} Overall, these results suggest that degree-degree correlations reduce the performance of the reconstruction process. Our intuition about those results is the following. In the case of negatively correlated nodes degree it can happen that if $k^{(1)}$ is small then $k^{(2)}$ is large and as well the opposite case, but then the multigraph will have, on average, constant nodes degree, being each degree the sum of a large and a small number, hence the matrix elements $F_{ij}$ given by Eq.~\eqref{eq:Fij} will be quite similar each other rendering difficult to solve the least-square problem. On the other hand if $G_1$ and $G_2$ have positive nodes degree correlations, then the multigraph will have nodes with very high degree, associated to large $k^{(1)}$ and $k^{(2)}$, together with nodes with very low degree, associated to small $k^{(1)}$ and $k^{(2)}$, the matrix elements $F_{ij}$ will thus be quite localized and the least-square problem easier to be solved. This could thus explain the dependence of the relative error on $\langle k_2\rangle$; on the other hand we already observed that $\langle k_2^2\rangle$ is more sensitive to the reconstruction process and thus deviations from the assumption of independence among the distributions of $k^{(1)}$ and $k^{(2)}$ manifest into a large relative error for any non zero $r$.

	\clearpage

\end{document}